\documentclass[manuscript]{acmart}

\AtBeginDocument{%
  \providecommand\BibTeX{{%
    \normalfont B\kern-0.5em{\scshape i\kern-0.25em b}\kern-0.8em\TeX}}}

\copyrightyear{2021} 
\acmYear{2021} 
\setcopyright{none}\acmConference[IUI '21]{26th International Conference on Intelligent User Interfaces}{April 14--17, 2021}{College Station, TX, USA}
\acmDOI{10.1145/3397481.3450656}
\acmISBN{978-1-4503-8017-1/21/04}

\begin{document}

\title{Perfection Not Required? Human-AI Partnerships in Code Translation}


\author{Justin D. Weisz}
\affiliation{
    \institution{IBM Research AI}
    \country{USA}
}
\email{jweisz@us.ibm.com}

\author{Michael Muller}
\affiliation{
    \institution{IBM Research AI}
    \country{USA}
}
\email{michael_muller@us.ibm.com}

\author{Stephanie Houde}
\affiliation{
    \institution{IBM Research AI}
    \country{USA}
}
\email{Stephanie.Houde@ibm.com}

\author{John Richards}
\affiliation{
    \institution{IBM Research AI}
    \country{USA}
}
\email{ajtr@us.ibm.com}

\author{Steven I. Ross}
\affiliation{
    \institution{IBM Research AI}
    \country{USA}
}
\email{steven_ross@us.ibm.com}

\author{Fernando Martinez}
\affiliation{
    \institution{IBM Argentina}
    \country{Argentina}
}
\email{martferc@ar.ibm.com}

\author{Mayank Agarwal}
\affiliation{
    \institution{IBM Research AI}
    \country{USA}
}
\email{Mayank.Agarwal@ibm.com}

\author{Kartik Talamadupula}
\affiliation{
    \institution{IBM Research AI}
    \country{USA}
}
\email{krtalamad@us.ibm.com}

\renewcommand{\shortauthors}{Weisz et al.}


\begin{abstract}
    Generative models have become adept at producing artifacts such as images, videos, and prose at human-like levels of proficiency. New generative techniques, such as unsupervised neural machine translation (NMT), have recently been applied to the task of generating source code, translating it from one programming language to another. The artifacts produced in this way may contain imperfections, such as compilation or logical errors. We examine the extent to which software engineers would tolerate such imperfections and explore ways to aid the detection and correction of those errors. Using a design scenario approach, we interviewed 11 software engineers to understand their reactions to the use of an NMT model in the context of application modernization, focusing on the task of translating source code from one language to another. Our three-stage scenario sparked discussions about the utility and desirability of working with an imperfect AI system, how acceptance of that system's outputs would be established, and future opportunities for generative AI in application modernization. Our study highlights how UI features such as confidence highlighting and alternate translations help software engineers work with and better understand generative NMT models.
\end{abstract}



\maketitle

\section{Introduction}
Recent advances in deep generative models have enabled them to produce content of a quality indistinguishable from that produced by a human being. These models have been rapidly adopted in many domains. For example, they have been used in creative domains to create or alter media such as images, videos, and artwork~\cite{bau2020semantic, li2019abstract, weber2020draw, liu2020generative, isola2017image}, music~\cite{yacht2019, huang2019bach, louie2020novice}, and writing~\cite{clark2018creative, gero2019metaphoria, roemmele2018automated}. They have also been used to produce artifacts such as novel furniture designs~\cite{autodesk2016elbo}, packaging designs~\cite{nutella2017}, aircraft compartment partitions~\cite{autodesk2016airbus}, and molecules~\cite{das2018pepcvae, chen2020machine, wei2019exploring}.

Recently, generative techniques have been applied to the realm of software engineering. Leveraging the naturalness hypothesis~\cite{allamanis2018survey} -- code is a form of human communication with similar statistical properties as natural languages -- new progress in neural machine translation (NMT) has demonstrated how unsupervised techniques can be used to train models that transform source code from one programming language to another~\cite{roziere2020unsupervised, chen2018tree}. However, code is unique compared to other kinds of natural languages: code is much more brittle, and swapping even a few characters or tokens can completely change its meaning or effect. In addition, code demands a certain level of correctness: code either compiles or does not, and it is either correct or contains bugs such as logic errors, security flaws, or race conditions. Given the probabilistic nature of deep learning models, we posit that there will always be some amount of noise in their output; correspondingly, deep generative models that produce code as output may exhibit some amount of deviation from a programmer's intentions. Indeed, while state-of-the-art NMT models -- e.g. \citet{tufano2019empirical}'s work on learning bug-fixing patches and TransCoder~\cite{roziere2020unsupervised} -- are often capable of producing high-quality source code, the code that they generate may not \emph{always} be error-free\footnote{In their evaluation of Java to Python translation, \citet{roziere2020unsupervised} report that 68.7\% of their tests contained a correct translation amongst the top 25 translation hypotheses, and the top choice was correct in only 35\% of cases.}.

Our research focuses on understanding the extent to which generative models are still \emph{useful} to human stakeholders despite their potential to produce imperfect output. Would people desire to work with a generative AI system \emph{knowing} that it might produce flawed output, or would they rather do the work themselves?  Are there ways that we can leverage generative model state to aid in the detection and correction of errors that do occur?  What characteristics of a user experience are important for facilitating acceptance of a deep generative model and its outputs?

We address these questions by considering the use of an NMT model within the context of application modernization~\cite{prolifics2020application, moore2018options, ibm2019application}. This context is both timely and relevant: timely, because many organizations are currently faced with the problem of how to migrate complex, legacy applications to the cloud when they do not necessarily have the resources or expertise to do so~\cite{gulla2019application, guerin2019legacy}; and relevant, because NMT models can specifically help in the process of translating from legacy languages or application frameworks (e.g. COBOL, J2EE) into modern ones (e.g. Python, Go).

To explore the utility of NMT models, we conducted a series of scenario-based design interviews with 11 professional software engineers who work across a variety of technology areas. Through a thematic analysis~\cite{clarke2015thematic, braun2006using}, we identified four themes when considering the design of user experiences that leverage generative models in software engineering: acceptance through verification, human-AI patterns of interaction, the utility of imperfect AI, and future opportunities for generative AI in application modernization. Our work makes three important contributions:

\begin{enumerate}
    \item We identify that although software engineers had concerns about the quality of the code produced by an NMT model, their concerns were tempered by the fact that integration of any NMT-produced code would follow existing best practices in software engineering (e.g. code reviews \& testing), thus ensuring that any potentially-flawed code would be subject to human review and revision before its inclusion in a product or service. Further, having interactive UX features such as confidence highlighting and alternate translations helped software engineers better understand how the NMT model produced its output, and even find bugs in its translations.
    \item We highlight how humans and AI systems both have unique and important roles to play in the context of application modernization, and discuss new opportunities for generative technologies to be applied across the application modernization lifecycle.
    \item We motivate the need to conduct further investigations into the design of human-AI partnerships that result in better outcomes than what could be accomplished by human or AI effort alone.
\end{enumerate}

\section{Related Work}
We discuss three areas relevant to our work: recent advances in the use of AI techniques, and specifically deep generative models, in software engineering; studies of the utility of imperfect AI; and studies of human-AI co-creation with generative AI.

\subsection{AI Techniques for Software Engineering}
In recent years, several efforts have focused on the use of AI and machine learning techniques for various tasks related to software engineering, including code completion~\cite{hindle2012naturalness,raychev2014code, bruch2009learning,svyatkovskiy2020intellicode}, code classification~\cite{MouLZWJ16, jayasundara2019treecaps}, API recommendation~\cite{gu2016continuous,BuiYJ19}, 
variable and method naming~\cite{allamanis2016convolutional,alon2019code2vec}, type inference~\cite{hellendoorn2018deep,wei2020lambdanet}, bug detection and repair~\cite{ray2016naturalness, pradel2018deepbugs,vasic2019neural, dinella2019hoppity,white2019sorting,hellendoorn2020global}, comment description and generation~\cite{moreno2013automatic,iyer2016summarizing,scalabrino2017automatically,hu2018deep,wan2018improving,alon2018code2seq}, code change summarization~\cite{moreno2014automatic}, and code clone detection~\cite{white2016deep}. A significant portion of this work is recounted in Allamanis et al.'s survey of the area~\cite{allamanis2018survey}. The emergence of generative AI techniques for natural language generation, such as GPT-2~\cite{radford2019language} and GPT-3~\cite{brown2020language}, have also been reflected in code-centric use cases: \citet{brockschmidt2018generative} 
proposed generative models for source code, and \citet{tufano2019empirical} 
used generative models for learning patches for bug fixes. In this paper, we focus on TransCoder~\cite{roziere2020unsupervised}, an unsupervised neural machine translation (NMT) model that transforms source code from one programming language to another.

\subsection{Imperfect AI}
Several studies have examined the notion of an imperfect AI system. \citet{kocielnik2019will} examined the effect of giving people control over the types of errors made by an AI meeting request detection system, either by avoiding false positives or false negatives. They found that, even when the system was only 50\% accurate (i.e. randomly guessing), users who expected a reduction in the false positive rate had a lower perception of accuracy and lower acceptance of the system than users who expected a reduction in the false negative rate. \citet{dzindolet2003role} discovered that once people observed an automated system make errors, their distrust in the system increased unless an explanation was provided regarding why the system might make an error; however, these explanations also increased trust and reliance on the system even when unwarranted, signaling the importance and difficulty of providing explanations that help people appropriately calibrate their trust. \citet{ehrlich2011taking} found a similar result, that giving a good reason for a poor recommendation made people more likely to accept that poor recommendation. \citet{zhang2020effect} and \citet{antifakos2005towards} both discuss how confidence scores can help calibrate people's trust in an AI model, and \citet{verame2016effect} show how such scores encourage use of an autonomous system. Our work examines a different aspect of imperfect AI by focusing on a co-creation scenario that has a lower tolerance for error (e.g. code ought to compile and be free of errors). Motivated by these studies, we also consider the effect that confidence and translation alternatives (which have explanatory power, discussed in Section~\ref{sec:ai-mechanics}) have on the utility and acceptance of a generative AI system.

\begin{figure*}[htp]
    \includegraphics[scale=0.195]{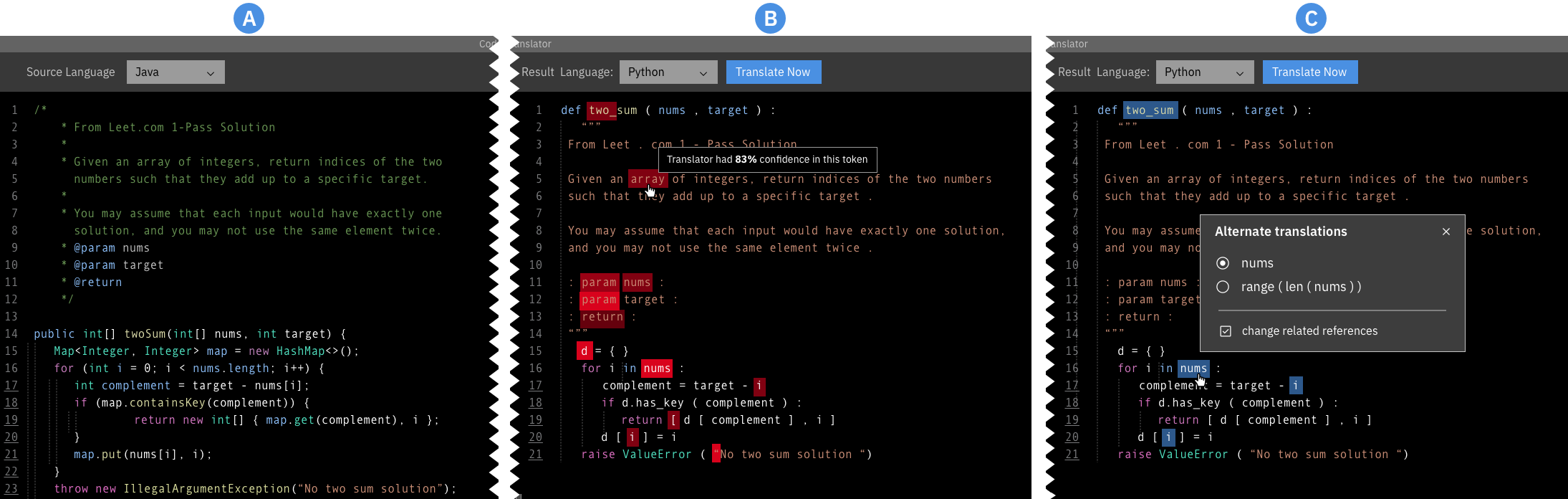}
    \centering
    \caption{UX Variants. Input Java source code was shown in the left pane (A). Clicking ``Translate Now'' translated the code into Python in the right pane, in one of three variants: code translation only (not shown), code translation with low-confidence tokens highlighted in red (B), or code translation with alternate code translations displayed in pop-up menus (C).}
    \label{fig:ide}
\end{figure*}

\subsection{Co-Creation with Generative AI}
Given the imperfect output of state-of-the-art NMT models, we posit that such systems will act in concert with human software engineers as a collaborative partner or teammate. The question of how to design effective interactions between human and machine teammates is of great interest to the IUI community. \citet{seeber2020machines} recently outlined a set of research questions spanning many aspects of human-AI collaboration, such as how tasks should be allocated across both parties and how trust could be established with a machine teammate. We find their question of how to determine an effective division of labor between humans and machines to be particularly salient as it is critical to understand what factors lead to productive human-AI partnerships.

Much of the recent focus of the generative AI community has been on developing new capabilities and exploring their applications in creative domains such as art~\cite{fan2019collabdraw, karimi2020creative, oh2018lead, weber2020draw}, photography~\cite{bau2020semantic}, music~\cite{huang2019bach, louie2020novice}, video games~\cite{guzdial2019friend}, and literature~\cite{clark2018creative}, as well as scientific domains such as drug discovery~\cite{das2018pepcvae}, materials discovery~\cite{yuan2020molecular}, and software engineering~\cite{mou2015end}. Some of this work has examined the extent to which deep generative models provide humans with augmented capabilities, addressing the question of the extent to which the human-AI team produces outcomes better than those produced by either party alone (e.g.~\cite{louie2020novice, guzdial2019friend, oh2018lead, fan2019collabdraw, weber2020draw, zhao2018compensation}). For example, \citet{louie2020novice} found that people working with a generative model to produce musical compositions felt a greater sense of self-efficacy and ownership of their compositions. \citet{weber2020draw} found that artwork that was restored with the aid of a generative model was rated as preferable to artwork restored by a human alone.

Despite identifying favorable outcomes of human-AI partnerships, we note that these outcomes are all subjective: the quality of the generated output lies in the perceptions of the people using the system. Less attention has been given to scenarios with objective outcomes, such as in software engineering where correct code is of prime importance. Given that the probabilistic nature of generative models implies the potential for some amount of error in their outputs, we seek to understand the extent to which they may be useful in scenarios that have an objective bar of quality.

\section{Design Scenario}
We developed an exploratory design scenario in order to engage software engineers in a discussion about the role of generative AI in application modernization. The use of scenarios to elicit feedback on future UX design has a long history in HCI (e.g.~\cite{carroll1995scenario, rosson2009scenario}), and recently this method has been applied to the design of AI systems as well (e.g.~\cite{avin2020exploring, liao2020questioning, wolf2019explainability}). As discussed by \citet[p. 252]{wolf2019explainability}, ``scenario-based design [is] a method that anticipates and leverages scenarios of \emph{possible use} early on in system development.'' (italics in original). We created a series of three progressively-enhancing UX variants within the scenario that illustrated different ways that a software engineer might interact with an NMT model to translate source code from one language to another. 

The UX variants were designed to resemble a simple programming environment with two code panes: a ``source'' pane on the left for input code, and a ``target'' pane on the right for the code output by the NMT model. Each variant appeared as a partially-functional prototype, through which participants could explore features of interest. Figure~\ref{fig:ide}A shows the base user interaction: a Java function\footnote{The code used in our scenario was drawn from \url{https://leetcode.com} and represents a solution to a non-trivial coding problem.} translated into Python using a pre-trained TransCoder model~\cite{roziere2020unsupervised}. Figure~\ref{fig:ide}B shows the first enhancement in which low-confidence tokens in the translated code are highlighted in red. Token-level confidences were extracted from TransCoder and presented as tooltips that displayed on mouse hover. Figure~\ref{fig:ide}C shows the second enhancement in which alternate translations were provided for some of the low-confidence regions. The alternate translations were obtained from TransCoder, which can generate multiple translations using beam search decoding~\cite{roziere2020unsupervised}. The UI also contained a feature to make relevant downstream edits when selecting an alternate translation (e.g. renaming all instances of a variable), as the selection of an alternate often carried consequences for other parts of the code.

Our scenario used real output from TransCoder that contained flaws. Although the translated code was syntactically correct, there was a logic error in the translation of a loop that resulted in wrong values being returned. There was also an issue with our reconstitution of tokenized code resulting in a discrepancy in an exception message string. These flaws provided a degree of realism to our study; several participants actually discovered these errors during the study, leading to discussions on the utility of imperfect AI.

\section{Method}
We conducted an interview study with software engineers in which we used the design scenario to spark discussions around the role of generative AI in application modernization. As the three UX variants built conceptually on top of each other, they were shown in the same order to each participant (A, B, C). The interviews lasted one hour and consisted of three phases. First, we discussed the participant's background, their level of experience with both Java and Python, and their experiences in modernizing legacy applications. If participants recalled highly-relevant experiences, we delved into more detail about their modernization project, how it was organized, what their role was, and what specific pain points existed in their work. Next, we showed each UX variant and allowed them to explore the interface while thinking aloud to capture their general impressions. Participants were also specifically asked what they did and did not like about the interface and how they might improve it. Participants were informed that the translation results were produced by an AI system and that the translations ``may be imperfect.'' After reviewing all three UX variants, we asked participants to brainstorm with us on broader applications of generative AI in application modernization. As a seed to the brainstorm, we showed three examples of how generative AI might provide unique value: translating complex algorithms, translating large codebases, and converting 3rd party library usage across languages. These ideas were generated through our pilot testing.

Each interview was attended by at least three of the authors. One author led each interview, guiding participants through each phase and asking all interview questions. The two other authors took running notes in a shared online document and occasionally stepped in to ask clarifying questions, probe deeper on topics of interest, or provide technical explanations for how the translation algorithm worked.

\subsection{Participants}
We recruited 11 full-time software engineers within our organization, an international information technology company. Participants were identified using an internal expertise-location tool that allowed us to find people in a software engineering role with experience in both Java and Python. Participants represented a diverse range of roles and technical backgrounds: Research Scientist in AI (P1); Software Engineer in AI (P0, P2, P6), Infrastructure Security (P5), and Mainframe (P9); Senior Software Engineer in Front-end Development (P10), Cloud (P3), and Healthcare (P7); and Software Architect in AI (P4) and Cloud (P8). Three participants presented as women; the other eight presented as men.

\subsection{Variant-Specific Questions}
For each UX variant, we asked participants about what they liked and disliked and how they might improve the design. We also asked specific questions for each variant. For Figure~\ref{fig:ide}A, we asked participants whether it was important to them to understand how the AI model produced the translation, and whether the AI is ``doing the right thing in modernizing a legacy app'' by translating code from one language to another, or whether it should ``do something else.'' For Figure~\ref{fig:ide}B, we asked participants if they understood why the highlighted tokens were highlighted, and what other information could help them better understand the AI's confidence. For Figure~\ref{fig:ide}C, we asked participants if they understood why the AI produced the alternatives it produced, and what other information could help them better understand how those alternatives were generated.

\section{Results}
We conducted a thematic analysis of our data to identify important ideas and themes from our interviews. Thematic analysis has been used for many kinds of qualitative analysis work in HCI (e.g.~\cite{macarthur2019makers}). We followed the process described by Braun and Clarke~\cite{clarke2015thematic, braun2006using} in which researchers familiarize themselves with their data, generate and group codes to identify higher-level themes, and iteratively refine codes and themes through collaborative discussion.

Our interviews generated a wealth of material: 11 hours of recorded videos, approximately 63 pages of notes, and a corpus of approximately 400 pages of interview transcripts containing about 89k words. For each interview, at least two authors reviewed each of the transcripts and notes and independently identified interesting observations. Initial descriptive codes were recorded in informal notes on these transcripts, which were then collaboratively reviewed by all authors. Next, we developed more integrative codes (similar to grounded theory axial codes) using a shared online document and visual collaboration tools. While there are no agreed guidelines about required sample sizes for this kind of analysis~\cite{morse1995significance}, we followed commonly-accepted practices to identify saturation (e.g. when we stopped identifying new codes in our interview transcripts)~\cite{guest2006many, majid2018achieving}.

We begin by highlighting the difficulties faced by our software engineers in modernizing legacy applications, motivating the need for AI support. We then discuss the four major themes we identified in our data: acceptance through verification, human-AI patterns of interaction, the utility of imperfect AI, and future opportunities for generative AI in application modernization.

\subsubsection{Challenges of Modernizing Legacy Applications}
\label{sec:challenges}
P4 described application modernization as ``a rats' nest [of] very impossible, undocumented, uncommitted, lots of dead unit tests that hadn't been run [in] year[s] kind of situation,'' and complained that ``there was an architecture at some point... then you had other folks come along and write additional pieces that... kind of go their own way.'' Many of these difficulties were due to decisions made by people who were no longer present in the team or in the company. P9 explained, ``I work on an application which was written before I was born,'' and P4 spoke of ``three generations of engineers that don't work here anymore.'' P7 described how legacy application specifications usually lived in the ``gray matter storage in other people.'' All participants who were involved in an application modernization project expressed some desire for help, and our focus on the code translation use case positively resonated with them.

\subsection{Acceptance Through Verification, Not Understanding}
\label{sec:acceptance-through-verification}
Many participants discussed the issue of trust in software engineering and how practices such as testing and code review help establish trust amongst human teams of software engineers. Participants expressed similar desires to establish trust with the code translation tool.

\begin{quote}
    \textit{``If I’m going to put my production systems functionality behind this, I really want some degree of trust that this thing is robust... Every piece of code that we put into production has been reviewed by at least one other person. And so we try hard to basically build that trust through review and familiarity in humans... we’d need basically the same level of trust for a system like this.'' (P4)}
\end{quote}

We recognize that trust is a complex, multi-faceted construct that has been described as being an attitude~\cite{rotter1967new}, an intention~\cite{mayer1995integrative}, and a behavior~\cite{adams2003trust}. In this paper, we adopt the AI-focused attitudinal view espoused by ~\citet{madsen2000measuring}, that trust is ``the extent to which a user is confident in, and willing to act on the basis of, the recommendations, actions, and decisions of an artificially intelligent decision aid.'' Thus, in our discussions with participants around the topic of trust, our emphasis was on understanding their attitudes toward acceptance of AI-generated code through their expressions of desire to incorporate (or not incorporate) such code in their work -- i.e. their ``willing[ness] to act,'' rather than their past behavior of having acted, as no participants had previously encountered a code-generating AI system.

\subsubsection{Acceptance Through Verification}
\label{sec:acceptance-through-verification}
In discussing whether they would accept code produced by the NMT model, participants felt that they would treat it in the same fashion that they would treat the output of their fellow software developers: by reviewing and testing it. P2 felt he would ``be happy with [the NMT model]'' as long as ``I get the correct output. All I would be concerned with is... if I had a five, six example test case, if they're producing the same output.'' P0 took a more extreme view of reviewing any code -- human or NMT-produced -- before accepting it, asserting, ``My developer attitude is `trust no one.' I mean, if someone hands me a piece of code, I don’t trust it. I test it.''

\subsubsection{Understanding AI Mechanics}
\label{sec:ai-mechanics}
Some participants had questions about the underlying mechanics of the NMT model, as they were ``curious how it works'' (P5) and ``would like to know what it's thinking'' (P10). However, participants ultimately felt that understanding how the NMT model worked was not a prerequisite for being able to put that model to productive use. When asked whether it was important to understand the mechanism of operation, P7 replied, ``Personally? I don’t care.'' P3 expressed the same sentiment, ``I guess I'm a little curious how it works, but maybe I don't wanna know [how it really works].'' P6 touched on the importance of evaluating the model based on the quality of its output: ``I don't really care [how it works]. If it's magic, I'm very happy with it.'' P1 expressed the same sentiment, citing efficiency gains as a reason for accepting the model's output without understanding how it was produced.

\begin{quote}
    \textit{``As a user, you... accept what's going on here as a useful and correct process and you probably do not even care how it's done because it will speed up your work... and at the end of the day, you'll say, `okay, I'm done.''' (P1)}
\end{quote}

P9 offered a more nuanced view on who \emph{would} need to have an understanding of how the AI model operated. He described a situation in which individual contributors working on a code conversion task ``don’t have to understand how the background of the application of the AI works, [because] they have to do the conversion [task].'' However, the architects overseeing the entire project ``have to know how this AI is working'' because they are the ones responsible for ``making this [code translation] project [happen].''

Despite the feelings that having an understanding of the NMT model's operation wasn't important, we observed that having such understanding does have benefits. For example, P6 described how understanding the model's operation might help him achieve better outcomes: ``I might want to have the option to look into what the AI is doing just so I can interpret or prepare my code better for the translation.''

Understanding the model's operation could also eliminate some of the confusion caused by the confidence highlights in Figure~\ref{fig:ide}B. Participants weren't told that it was a deep learning-based NMT model that produced the translation, so many assumed the translation was produced by a rule-based system. As asserted by P7, ``You have rules that are driving this confidence [feature].'' We observed that discrepancies between our participants' mental models and the actual operation of the NMT model was a source of confusion. P5 questioned, ``Why is it highlighting so many of these comments?'' P1 was surprised that the model was not confident in some of the punctuation tokens: ``Why is the left bracket after the return... why is that not so confident... it's not even an identifier... the double quote in that string there, that's a funny one too.'' P3 expected that the ``translator would be able to define this def function signature line with the function name correct,'' but expressed confusion when the translator wasn't ``as confident as it could be.'' Similarly, P0 struggled to understand why the NMT model would not be confident in a line of code that was syntactically correct.

\begin{quote}
    \textit{``That just confuses me because this line of code is syntactically correct. I would write that line of code with 100\% confidence. I'm confused about why the translator would be confused about that assignment.'' (P0)}
\end{quote}

This discrepancy in confidence, in which participants were more confident in their own abilities to translate the code than the NMT model, caused many participants to desire additional insight into \emph{why} the NMT model was not confident. P2 said, ``[If it's] 42\% confident about this translation, maybe... [there could be] a link that you could click on to see why it is not confident.'' P8 similarly suggested, ``An explanation of why the [model had] low confidence... would be helpful.'' Thus, although participants generally said they ``don't care'' (P7) about knowing how the translation was produced, explanations of the model's mechanics may be helpful for avoiding confusion caused by confidence discrepancies.

One way of producing such explanations is via the alternate translations shown in Figure~\ref{fig:ide}C. The presence of the alternate translations seemed to provide an implicit explanation that helped participants better understand the model. P7 described how one of the alternatives more explicitly demonstrated the code's functionality: ``The first [option] is more Pythonic because it's taking advantage of the built-in functionality of an array... however this [second option] is far more explicit... this is what's happening behind the scenes.'' P4 also gained insight by comparing two alternate translations: ``This is clearly... toggling between those two styles of iteration... it's interesting to see what the AI is... thinking between these two different programming constructs.''

\subsubsection{Feedback \& Learning}
Participants identified the importance of being able to give ``feedback'' (P4), such as by ``unflag[ing]'' (P7) low-confidence tokens identified by the NMT model. P4 specifically called out how code reviews are used to give people feedback to improve their skills and how such reviews could be used to improve the NMT model.

\begin{quote}
    \textit{``With humans, you can teach through review...if somebody is continuously making these kind of mistakes, and we're catching them in review, that person is going to get annoyed by a million comments... you could probably have some sort of reinforcement or sort of iterative learning in your model that would teach it.'' (P4)}
\end{quote}

P6 further imagined that ``I’m just going to rewrite this and through my action the AI can learn.'' Similarly, P10 would ``make my edits, and... that would help train the model.''

\subsection{Human-AI Patterns of Interaction}
Participants responded positively to the confidence highlights (Figure~\ref{fig:ide}B) and the alternate translations (Figure~\ref{fig:ide}C) and felt that both views were helpful. Their comments provided insight into how these features could be used to drive human-AI workflows and how human actions could provide feedback to improve the NMT model. As discussed by \citet{seeber2020machines}, the allocation of tasks in a human-AI collaboration need not be symmetric. P4 summarized his ideal interplay between human and machine: ``Focus on letting the machine do the stuff that's really easy and letting a human help out with the parts that are less easy.''

\subsubsection{Translation Meta-Information Guides Human Task Management}
We initially thought that the translation would produce only \textit{content} in the form of the translated code. We learned that the low-confidence highlighting (Figure \ref{fig:ide}B) could also help developers to organize their work. P10 crystallized this notion by speaking of a ``checklist,'' and P7 desired the ability to turn off the highlights after reviewing each one. P4 emphasized the utility of this approach: ``Having something to draw my attention to places where it's more likely that there’s an error would be super helpful.'' Similarly, P6 said, ``If it's trying to highlight that it's really not sure that it's done a good job here then I should pay attention to this piece and see if it makes sense.'' P5 pointed out that, ``If you used this [confidence highlighting] on a big code base... you could know where to look, where there might be problems.'' Thus, the meta-information provided by the existence of a highlight can help ``steer'' human activity, complementing Louie et al.'s notion of how humans may ``steer'' the output of a generative model~\cite{louie2020novice}. Many participants also pointed out how edits to low-confidence translated code could serve as feedback to the NMT model to improve its training.

\subsubsection{Alternate Translations as Explanations and Quick Fixes}
Many participants were initially skeptical of the accuracy and value of the confidence highlights (P1, P2, P3, P5, P7, P9). Once they saw the alternate translations, they began to understand \textit{why} the NMT model had highlighted certain tokens as low-confidence: precisely \textit{because} there were alternate ways of translating those tokens. Thus, the alternate translations served as explanations for the highlights.

In addition, the alternate translations had additional explanatory power in helping participants understand the nature of the underlying code. Both P1 and P3 were able to identify a logical error in the translated code \emph{because} they reviewed the alternate translations.

\begin{quote}
    \textit{``It would be different if the i as an index... let's see... is it used? Oh actually it is used... Now there's a mistake in the code.. The statement d [ i ] = i, is wrong [but] the other [alternate translation] is correct.'' (P1)}
\end{quote}

Participants also appreciated the feature of the alternate translation UI in which downstream edits could be made when selecting an alternate translation (e.g. because selecting an alternate may have renamed a variable or changed a loop index). P7 commented, ``the AI’s letting me know... it's best if you change it in multiple locations in order to be consistent.'' In this way, the NMT model can work collaboratively with the human: the model proposes multiple solutions, the human chooses one, and if necessary, the model follows-through to make a series of changes that are consistent with one another. Multiple participants pointed out how this interaction pattern would enable a feedback loop to improve the NMT model (P4, P6, P7).

\subsection{Utility of Imperfect AI}
\label{sec:utility}
Software developers typically work with deterministic tools and a great deal of work goes into ensuring that these tools are highly reliable~\cite{chen2011compiler}. The possibility that a software problem would be introduced by a compilation process, for example, would not be considered under ordinary circumstances. Thus, we questioned whether we would encounter resistance to a tool that was \emph{not} guaranteed to produce correct results. To probe at this idea, we selected a code sample with an incorrect translation in order to spark discussion on the utility of erroneous output.

In some cases, although not an explicit part of the study, participants actually found the errors in the translation. For example, P4 found a string mismatch: ``There is an actual functional difference here on line 21 of the Python code where there's an extra space inserted into that error string.'' As previously mentioned, P1 discovered a logic error: ``Ah, see there, now there is a mistake in the code. The statement d [ i ] = i is wrong.''

Overall, many participants felt that even the erroneous output was desirable and that it would be something they would find useful to have. P5 directly asserted, ``Yes, the imperfect translations are useful.'' P2 described how the translations, even with small mistakes, were preferable to manually rewriting code.

\begin{quote}
    \textit{``Yeah I would [like it]. It makes it much more easier than to just rewrite the code or to hand type the code. Even if it had... little distortions, like extra spaces, which can be easy to fix... I would rather do that instead of rewriting everything from scratch.'' (P2)}
\end{quote}

Despite the positive reception of potentially error-prone code, there did seem to be a threshold regarding how many errors would be acceptable. As summarized by P6, ``If it's egregiously wrong then no, I'd rather do it myself. But as long as it's giving me a head start of any sort, I'll take it.'' P3 described how there may be individual differences in the value people derive from NMT translations.

\begin{quote}
    \textit{``I think for someone to start using a tool like this, it would need to prove that it's adding some value... It probably varies from person to person. Some people won't be satisfied with anything you give them, and others will be satisfied with if it can be 10\% effective in doing translations perhaps. And most people fall somewhere in the middle there.'' (P3)}
\end{quote}

Participants also expressed mixed feelings about their tolerance for easy-to-spot or easy-to-fix errors. On one hand, P5 felt that ``I think it would still be worth using the tool if the only problems were small things like that, little nitpickings.'' On the other, P0 expressed that errors that weren't related to the ``central problem'' might interrupt his flow and become distracting if they occurred too often: ``If I'm iterating and I have to stop and fix something else that’s not germane to the central problem, that can become annoying if it's repetitive.''

For large codebases, the downside of imperfectly-translated code may be less than the upside of simply having automatically-produced translations. P3 discussed how the NMT model could fit into a workflow for translating a large codebase.

\begin{quote}
    \textit{``A lot of us in the software development field, we sometimes have the feeling that for a given problem set, we have to solve the entire problem. But, that's not necessarily true... if you can help someone do part of the task... maybe it will be more effective at smaller, less complex code fragments. Then, [as] you get feedback from users, you build into it more intelligence, then maybe it becomes more effective on larger, more complex pieces of code.'' (P3)}
\end{quote}

Thus, although many participants expressed a willingness to accept imperfectly-translated code as a starting point, acceptance may depend on how many errors the translated code contains as well as the nature of those errors (e.g. if they are easy to spot and fix, or if they take attention away from the ``central problem''). P3 succinctly summarized this point: ``Even if some things were incorrect in the [translation], I think there's still a potential for the tool saving the developers some time... it doesn't have to be perfect and it doesn't have to necessarily be complete for a tool like this to be effective.''

\subsection{Beyond our Design Scenario: Opportunities for Generative AI Across the Modernization Lifecycle}
Participants described the lifecycle of modernizing applications as consisting of three general phases: (1) creating an understanding of the legacy application and its architecture by reviewing code and documentation; (2) performing the migration work, which may include translating and/or refactoring the code; and (3) reviewing and testing the migrated code. Throughout our interviews, participants made a number of suggestions for how generative AI models, as well as the user interfaces that incorporate them, could support these activities.

\subsubsection{Creating Understanding}
As described in Section~\ref{sec:challenges}, one of the main challenges in migrating legacy applications is due to the fact that code and code architectures are ``undocumented,'' (P4) with specifications sometimes living in the ``gray matter storage'' (P7) of ``engineers that don't work here anymore'' (P4). P6 described how generative methods could help him form a clearer picture of system architecture: ``If AI could actually somehow... help me keep track of how the current function I’m looking at fits into the bigger picture... I think that's really beneficial.'' P4 described how generative methods could be used to ``reverse engineer the architecture programmatically'' and deal with ``dead'' codebases.

\begin{quote}
    \textit{``The place where I can see something like this being very helpful would be where you really have a truly dead codebase that's just a complete black box that nobody can, or wants, to understand... because for that you need understanding if the goal is to keep the darn thing limping while you transition to something else.'' (P4)}
\end{quote}

Translation could also help software engineers better understand codebases in languages with which they are less familiar: ``I've got a bunch of engineers that know Python, but they don't know Java. So, it might be easier to delve in and spelunk through... auto gen[erated] Python code than it would be to go back and spelunk the original Java.'' (P4).

Participants also described how generative methods could be used to fill in missing documentation to help others understand one's own code, to create an understanding of poorly-documented code, or to re-align documentation with code when drift has occurred~\cite{fluri2007code, tan2012tcomment}. P6 described how his team has strict requirements for writing documentation for every function because it is used to generate Swagger API documentation\footnote{https://swagger.io}. Interestingly, he would tolerate mistakes or errors in the generated documentation.

\begin{quote}
    \textit{P6: ``It's very tedious to write these comments... It really takes so much time... If the AI generated this for me I think that is gonna be really helpful.''} \\
    \textit{Interviewer: ``What if the AI was only able to generate part of it, or maybe sometimes made mistakes...Would you still want to use it... or would you rather just write [it] from scratch?''} \\
    \textit{P6: ``As long as it is giving me a head start of any sort I will take it because for me the overhead to write such comments for every function that I code is very high. As you can imagine in a practical system there is going to be hundreds of such functions... it’s a lot of human hours.''}
\end{quote}

Generating documentation from source is an active area of research in the machine learning community (e.g.~\cite{moreno2013automatic, iyer2016summarizing, hu2018deep, wan2018improving}) and our results highlight the importance of this functionality in user experiences, even when generated documentation may be imperfect.

\subsubsection{Performing Migration Work}
One helpful improvement to the code translation user interface would be to show a stronger visual correspondence between input and output source code. As described by P4, ``If I had [it] sort of laid out with a diff, I might just be able to parse it visually a little better.'' In addition, several participants noted the importance of coding styles (e.g. variable naming conventions, documentation standards) and the desire for output code to have a ``standard formatting'' (P5) because ``reading other people's code is much easier when you all adhere to the same standards.'' (P5). These comments call out that an AI teammate would be expected to follow the same coding conventions as other members of the team.

Many participants described how code translation could be used to take advantage of functionality defined in third party libraries, ``because the same libraries in Java don't exist in Python, or similar libraries exist but they're not... syntactically the same.'' (P0). P10 described, ``If you had a library that you know was in another language and you want to use it, but they don’t have it in your language, that’s great.'' P3 discussed how he had to manage libraries for the same set of APIs across multiple languages: ``We're up to nine different languages now total and and each one seems to have its own, like, name capitalization conventions and things like that.'' Generative techniques that are capable of converting code from one language to another, from one version of a language to another, or even from one version of an API or framework to another, would be helpful. P8 described how the code translation process wasn't just about converting from one language to another, but also about identifying opportunities to leverage language-specific features to create more optimal code: ``The whole point of using the latest version is to tap into the new features... in JDK 8 they have lambdas... it is much faster. If it could do something like this I would use it like hell!'' P9 emphasized how the alternate translations provided an educational opportunity while doing code modernization work because ``having new options makes it much more easy for me to explore newer methods in the newer language.''

Finally, P7 discussed how an AI partner might monitor his work and make proactive suggestions: ``It would be cool if the AI knew what I was trying to build, like a controller or view, and it would finish it for me. Like give me a shell of what I'm trying to do so I can fill in the details.''

\subsubsection{Reviewing \& Testing}
As discussed in Section~\ref{sec:acceptance-through-verification}, testing seems to be one of the primary ways that software engineers will accept and use AI-translated code: ``I'm very much a supporter of test-driven development, which typically means you write tests that validate behavior.'' (P3). Not surprisingly, unit test generation was a highly-requested feature for this stage of application modernization. This notion was best summarized by P4.

\begin{quote}
    \textit{``One thing that would be really interesting... especially if you are going from a strongly-typed language like Java to something like Python, would be to generate automated unit tests with coverage. That would certainly help probe at places where the AI got it wrong and build trust where it got it right.'' (P4)}
\end{quote}

Generating a ``whole slew of automated unit tests'' (P4) as well as ``more realistic test values'' (P3) ensures that ``the right hand side code not only runs and doesn't crash but produces expected output'' (P4). Indeed, the ML community has explored many methods for generating tests~\cite{taneja2008diffgen, thummalapenta2009mseqgen, fraser2014large} as well as generating data for tests~\cite{jeske2006synthetic, soltana2017synthetic}, and our findings reinforce the need to incorporate these techniques into mainstream application modernization tools.

\section{Discussion}

\subsection{Human-AI Partnerships}
Participants were eager to accept help from an imperfect AI assistant (e.g. as discussed by P2 \& P3). Overwhelmingly, they felt that a tool that could do code translation work within the application modernization context would be useful as it would help them get their own work done more quickly and effectively (e.g. as discussed by P3). Instead of viewing the review and correction of the translated code as additional effort, they felt that the skeletal framework provided by the NMT model would be a useful starting place that they could then adjust and correct as necessary, as they usually do with code produced by their colleagues (e.g. as discussed by P4). Additional support provided by the NMT model by way of confidence highlights and alternate translations were also seen as critical ways to help prioritize their review effort. We also discovered how this information possessed explanatory power, helping participants both understand the code's underlying intent and identify potential errors in the translation.

We note that our system only examined one kind of human-AI interaction pattern, in which the human provided code and the translator produced a translation. Although the system did allow for some subsequent actions by the human (e.g. hovering to see confidence scores or clicking to see alternate translations), our tool falls short of the tightly-coupled human-machine interactions envisioned by some of the informants in Seeber et al.'s study~\cite{seeber2020machines}. Participants did identify how to ``close the loop'' by using the selection of alternate translations to provide feedback to the NMT model to improve its training. However, our system was solely driven by human initiative. Future work is needed to investigate different kinds of interaction patterns, including different mixed-initiative patterns~\cite{shneiderman2020human, horvitz1999principles}, to understand their effect on a software engineer's productivity and capability. For example, we believe that there is a richer design space in which AI translations can be provided at a number of different granularities, from highly-automated, bulk translation of large codebases to highly-controlled, real-time translation of code as it is being written.

\subsection{Toward Explainable Generative AI}
While conventional software development tools are generally quite reliable, software engineers themselves are not. They recognize the human capacity to err in themselves and their colleagues, and thus have developed processes and ways of working in which these errors are expected: code reviews~\cite{bacchelli2013expectations, mcintosh2016empirical}, test-driven development~\cite{cordemans2014test, maximilien2003assessing}, and root-cause analysis~\cite{lehtinen2011problem} are all ways of finding, fixing, and learning from the mistakes that are bound to happen in any software engineering effort. As a result, participants felt that having an understanding of the mechanics of the NMT model was less important for accepting its output than just reviewing and testing that output. That said, participants who did understand that the NMT model was probabilistic, and not rule-based, in nature, were less confused by spurious confidence highlights. In addition, the explanatory power of alternate translations enabled participants to understand \emph{why} regions of the code were flagged as low confidence, helping them focus their effort toward reviewing those potentially problematic sections. Alternate translations also helped participants identify errors in the translation by showing them alternatives that the translator had considered.

Our results motivate the need for explainable generative AI. Techniques that help people understand the mechanics of generative models, as well as techniques that are able to generate specific rationales for why a given artifact was generated, could help people further adopt generative models in their own work practices. We recognize that some work has been done in this space, such as in visualizing how transformer models produce their output (e.g.~\cite{hoover2019exbert}). However, more work is needed to understand how such visualizations and other explanatory methods impact peoples' willingness to use generative models and their productivity in using them.

\subsection{Future Directions for Generative AI in Application Modernization}
Our study identified a number of areas that could be pursued for the use of generative AI in application modernization. One such area is in further developing intelligent user interfaces tailored specifically for code translation. For example, participants desired a stronger visual correspondence between the source and translated code to understand how one flowed from the other. In addition, task-oriented displays of the ``checklist'' mentioned by P10 could be used to help organize and track review efforts.

Incorporating feedback to the generative model from the edits made by software engineers is another fruitful research opportunity. NMT models such as the one we used are typically trained in an unsupervised manner, in part due to the lack of a corpus of parallel code examples across source and target languages. The process of using these models to perform an initial translation, followed by manual correction, could provide a corpus of correctly-translated code that would enable supervised training, resulting in further quality improvements to the translator. In addition, technical mechanisms that enable fine-grained human control over the translation process ought to be explored.

Participants identified many opportunities for generative models to aid the overall application modernization workflow of building understanding, performing the migration, and reviewing \& testing. Although the import of generative capabilities may seem to be about the production of an artifact itself, we found that they instead may also be a \emph{means} rather than just an \emph{end}. For example, the generation of documentation and architecture specifications, rather than being the desired end products themselves, are more of an intermediate output that supports a modernization effort by helping people understand an ``impossible, undocumented'' (P4) legacy application. Thus, we challenge the community to consider use cases for generative AI that involve improving the state of human understanding of a domain or artifact, rather than considering the only utility of generative models to be the output itself.

\section{Limitations}
Our study is a preliminary examination of how probabilistic generative models can be applied to a use case that requires an objective level of quality. Although we have discovered that our design scenario did demonstrate compelling user experiences, we caution that our results may not generalize to other domains or use cases beyond the code translation task we examined (e.g. using generative models for molecular discovery). In addition, our finding that acceptance of a model's output is established through verification, rather than explanation of the model's operation, is not meant to diminish the importance of further research and exploration into how generative models might explain themselves, their operation, and their limitations. Indeed, having an accurate mental model did help participants understand the model's confidence levels, and we hypothesize that generative AI user interfaces that provide salient details of a model's operation can help people form accurate mental models.

\section{Conclusion}
Generative AI techniques are enabling new forms of human-AI co-creation. Recent advances in deep generative models have enabled them to produce content with a proficiency often indistinguishable from that of humans. New unsupervised neural machine translation models have demonstrated the capability for generative AI to translate code from one programming language to another in the software engineering domain. We have examined how such a model would be received by software engineers in the context of modernizing legacy applications. Counter to expectations, software engineers felt a generative model that produced imperfect output -- such as code with compilation or logical errors -- would be a useful aid, albeit one subject to the same procedures of review and testing as code produced by their colleagues. Interface features such as confidence highlighting helped steer human attention toward potentially problematic areas, and alternate translations provided explanatory power for why a segment of translated code was rated as low confidence by the model; in addition, alternate translations provided clues toward the semantic meaning of the code and helped participants find logical errors in it. Future work should explore how different patterns of interaction and initiative, as well as capabilities beyond code translation, can enable effective human-AI partnerships with generative models that produce better outcomes than what could be accomplished by either party alone.

\bibliographystyle{ACM-Reference-Format}
\bibliography{refs}

\end{document}